\documentclass[11pt]{article}
\usepackage[french,english]{babel}
\usepackage{amsmath}
\usepackage{amsfonts}
\usepackage{amssymb}
\usepackage{graphicx}   
\usepackage{times}
\def\figskip{\vskip .5cm plus 3mm minus 2mm}
\begin{document}
%
\begin{titlepage}
June 12th 2007\hfill
\vskip 5.5cm
{\baselineskip 20pt
\begin{center}
{\bf
           MIXING ANGLES OF QUARKS AND LEPTONS AS AN OUTCOME \break
                 OF $\boldsymbol{SU(2)}$ HORIZONTAL SYMMETRIES
}
\end{center}
}
\vskip .2cm
\centerline{Q. Duret
    \footnote[1]{E-mail: duret@lpthe.jussieu.fr}
          \& B. Machet
     \footnote[2]{E-mail: machet@lpthe.jussieu.fr}
     \footnote[3]{\selectlanguage{french}{Member of \flqq Centre National
de la Recherche Scientifique\frqq}}
     }
\vskip 5mm
\centerline{{\em Laboratoire de Physique Th\'eorique et Hautes \'Energies}
     \footnote[4]{LPTHE tour 24-25, 5\raise 3pt \hbox{\tiny \`eme} \'etage,
          Universit\'e P. et M. Curie, BP 126, 4 place Jussieu,
          F-75252 Paris Cedex 05 (France)}
}
\centerline{\em Unit\'e Mixte de Recherche UMR 7589}
\centerline{\em Universit\'e Pierre et Marie Curie-Paris\,6 / CNRS /
Universit\'e Denis Diderot-Paris\,7}
\vskip 1cm
{\bf Abstract:} We show that all mixing angles are determined, within
experimental uncertainty, by a product of $SU(2)$ horizontal symmetries
intimately linked to the algebra of weak neutral currents.
This concerns: on one hand, the three quark mixing angles;
on the other hand, a neutrino-like pattern in which $\theta_{23}$ is maximal
 and $\tan (2\theta_{12})=2$.  The latter turns out to exactly satisfy the
``quark-lepton complementarity condition'' $\theta_c + \theta_{12}= \pi/4$.
Moreover, among all solutions, two  values for the third neutrino
mixing angle arise, which satisfy the bound $\sin^2(\theta_{13}) \leq 0.1$:
$\theta_{13} = \pm 5.7\,10^{-3}$ and $\theta_{13} = \pm 0.2717$.

\bigskip


\vfill
\vfill
\begin{center}
\includegraphics[height=1.8truecm,width=4truecm]{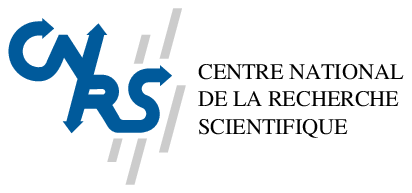}
\end{center}
\end{titlepage}
%
\section{Introduction}
\label{section:introduction}

Following the study of  neutral kaons  done in \cite{MaNoVi},
we have shown in \cite{DuretMachet1} and \cite{DuretMachet2} that:

$\ast$ in Quantum Field Theory (QFT), mixing matrices linking
flavour to mass eigenstates
 for non-degenerate coupled systems should never be parametrized as
unitary. Indeed, assuming that the effective renormalized quadratic
Lagrangian
is hermitian at any $q^2$ and that flavour eigenstates form an
orthonormal basis, different mass eigenstates, which correspond to
different values of $q^2$ (poles of the renormalized propagator)
 belong  to different orthonormal bases
\footnote{Since at any {\em given} $q^2$,
the set of eigenstates of the renormalized quadratic Lagrangian form
an orthonormal basis, the mixing matrix with all its elements evaluated at
this $q^2$ is unitary and
the unitarity of the theory is never jeopardized.\label{ftn:unitarity}};

$\ast$ when it is so, the properties of universality for diagonal neutral
currents and absence of flavor changing neutral currents (FCNC)
which are systematically implemented, for the
Standard Model (SM), in the space of flavour eigenstates, do not
automatically translate anymore into equivalent properties in the space of
mass eigenstates. In the case of two generations of fermions,
imposing them for mass eigenstates yields two types of solutions for
the mixing angles
\footnote{For two generations, one is led to introduce two mixing angles to
parametrize each $2 \times 2$ non-unitary mixing matrix.}
 of each doublet with identical electric charge:
Cabibbo-like solutions
\footnote{Cabibbo-like angles  can only be fixed by imposing
conditions on the {\em violation pattern} of the unitarity of the mixing
matrix in its  vicinity.} which reduce to a single
unconstrained mixing angle, and a set of discrete solutions, unnoticed in
the customary approach, including in particular the so-called maximal
mixing $\pi/4 \pm k\pi/2$;

$\ast$ for any of these solutions one recovers a unitary mixing matrix;
but, as said above, very small deviations are expected due to mass
splittings, which manifest themselves as a tiny departure from the exact
two previous conditions.
In particular, in the {\em neighborhood} of a Cabibbo-like solution,
these deviations  become of equal strength for a value of
the mixing angle extremely close  to the measured Cabibbo angle 
\begin{equation}
\tan (2\theta_c) = 1/2.
\end{equation}
This success was a encouragement to go further in this direction.
We present below the outcome of our investigation of neutral current
patterns  in the case of three generations of fermions.
In particular, we show that the requested scheme of unitarity violation
of the mixing matrices trivially interprets  in terms of a product of $SU(2)$
horizontal symmetries, implemented at the level of neutral weak currents.
Hence,  the values of all mixing angles, in the quark as
well as in the leptonic sector, are controlled by this symmetry.

The intricate system of trigonometric equations has been solved by
successive approximations, starting from configurations in
which $\theta_{13}$ is vanishing. We will see that this approximation,
obviously inspired by the patterns of mixing angles determined from
experimental measurements, turns out to be a very good one.
Indeed, we precisely show, without exhibiting  all the solutions of our
equations, that
the presently observed patterns of quarks as well as of neutrinos, do
fulfill our criterion.
While the three angles of the Cabibbo-Kobayashi-Maskawa (CKM)
 solution are ``Cabibbo-like'', the neutrino-like solution
\begin{eqnarray}
\tan (2\theta_{12}) &=& 2\  \Leftrightarrow\  \theta_{12}\  \approx\  31.7^o,\cr
\theta_{23} &=& \pi/4,\cr
 \theta_{13} &=& \pm 5.7\, 10^{-3}\ \text{or}\ \theta_{13}=\pm 0.2717
\label{eq:nupre}
\end{eqnarray}
is of a mixed type, where  $\theta_{23}$ is maximal
 while $\theta_{12}$ and $\theta_{13}$ are Cabibbo-like.

Two significant features in these results must be stressed. First,
the values for the third neutrino mixing angle
 $\theta_{13}$ given in (\ref{eq:nupre}) are predictions
which take into account present (loose) experimental constraints.
Only two possibilities survive: an
extremely small value $\theta_{13} \sim V_{ub} \sim$ a few
$10^{-3}$, and a rather ``large'' one, at the opposite side of the
allowed range. Secondly, our procedure yields in an exact, though quite
simple way, the well-known ``quark-lepton complementarity relation''
\cite{QLC} for 1-2 mixing:
\begin{equation}
\theta_{12} + \theta_c  = \pi/4,
\label{eq:qlc}
\end{equation}
where $\theta_{12}$ is the leptonic angle, and $\theta_c$ the
Cabibbo angle for quarks.

\section{Neutral currents of mass eigenstates and
$\boldsymbol{SU(2)}$ symmetries}
\label{section:general}

\subsection{The different basis of fermions}
\label{subsection:basis}

Three bases will appear throughout the paper:

$\ast$ flavour eigenstates, that we note
$(u_f, c_f, t_f)$ and $(d_f, s_f, b_f)$ for quarks, $(e_f, \mu_f, \tau_f)$
and $(\nu_{ef}, \nu_{\mu f}, \nu_{\tau f})$ for leptons;

$\ast$ mass eigenstates that we note $(u_m, c_m, t_m)$ and
$(d_m, s_m, b_m)$ for quarks, $(e_m, \mu_m, \tau_m)$
and $(\nu_{em}, \nu_{\mu m}, \nu_{\tau m})$ for leptons; they include 
in particular the charged leptons detected experimentally, since 
their identification proceeds  through the measurement
of their $charge/mass$ ratio in a magnetic field; 

$\ast$ the neutrinos that couple to mass
eigenstates of charged leptons in charged weak currents.
These are the usual "electronic'', ``muonic" and "$\tau$" 
neutrinos $\nu_e$, $\nu_\mu$, $\nu_\tau$ considered in SM textbooks
\cite{Vysotsky}: they are indeed identified by the outgoing charged
leptons that they produce through
charged weak currents, and the latter are precisely mass eigenstates
 (see above).  These states read  (see Appendix \ref{section:CC})

\begin{equation}
\left(\begin{array}{c} \nu_e \cr \nu_\mu \cr \nu_\tau \end{array}\right)
= K^\dagger_\ell 
\left(\begin{array}{c} \nu_{ef} \cr \nu_{\mu f} \cr \nu_{\tau f}
 \end{array}\right)
= (K^\dagger_\ell K_\nu)
\left(\begin{array}{c} \nu_{em} \cr \nu_{\mu m} \cr \nu_{\tau m}
 \end{array}\right),
\label{eq:elecnu}
\end{equation}
where $K_\ell$ and $K_\nu$ are the mixing matrices respectively of charged
leptons and of neutrinos ({\em i.e.} the matrices that connect their flavour
to their mass eigenstates).
Note that these neutrinos coincide with flavour eigenstates when the
mixing matrix {\em of charged leptons} is taken equal to unity
$K_\ell =1$, {\em i.e.} when the mass and flavour eigenstates
of charged leptons are aligned, which is often assumed  in the literature.

\subsection{Neutral currents  and $\boldsymbol{SU(2)}$ symmetry}
\label{subsection:NCMM}

The principle of the method is best explained in the case of two
generations. This in particular makes the link with our previous work
\cite{DuretMachet2}.

Let us  consider for example the $(d,s)$ channel
\footnote{``Channel $(i,j)$''
 corresponds to  two fermions $i$ and $j$ with identical electric charge;
for example, ``channel $(2,3)$'' corresponds to $(d,b)$,
 $(c,t)$, $(\mu^-, \tau^-)$ or $(\nu_\mu, \nu_\tau)$.}.
The corresponding neutral currents in the basis of mass eigenstates
are controlled by the
product $C^\dagger C$ of the  mixing matrix $C$ between $d$ and $s$ with its
hermitian conjugate (see \cite{DuretMachet1}). Requesting the absence of
non-diagonal currents and universality for diagonal currents (that we call
hereafter the ``unitarization'' conditions) selects two classes of mixing
angles \cite{DuretMachet1}:
a first class that we call ``Cabibbo-like'' which is unconstrained at this
level, and a second class made of a series of discrete
values, maximal or vanishing. As soon as $C$ departs from
unitarity, the  Lagrangian for neutral currents reads, in the basis of mass
eigenstates
\begin{equation}
{\cal L} \propto W_\mu^3 \left[\alpha\, \bar d_m \gamma^\mu_L d_m +
\beta\, \bar s_m \gamma^\mu_L
s_m + \delta\, \bar d_m \gamma^\mu_L s_m + \zeta\, \bar s_m \gamma^\mu_L d_m\right].
\label{eq:Lds}
\end{equation}
The condition that was shown in \cite{DuretMachet2} to fix the value of
the ``Cabibbo'' angle, {\em i.e.} that the universality of 
$\bar d_m \gamma^\mu_L d_m$ and $\bar s_m \gamma^\mu_L s_m$ currents is
violated with the same strength as the absence of $\bar d_m \gamma^\mu_L
s_m$ and $\bar s_m \gamma^\mu_L d_m$ currents, now simply reads
$\delta = \alpha -\beta = \zeta$,
which states that $\cal L$ in (\ref{eq:Lds}) is invariant by the $SU(2)$
symmetry which rotates  $d$ and $s$. Eq.~(\ref{eq:Lds}) indeed
trivially rewrites, then,
\begin{eqnarray}
{\cal L} &\propto& W_\mu^3 \left[
(\alpha+\beta) \frac{\bar d_m \gamma^\mu_L d_m + \bar s_m \gamma^\mu_L
s_m}{2}\right.\cr
&&\left. \hskip 1cm +\; (\alpha-\beta) \left(\frac{\bar d_m \gamma^\mu_L d_m -
\bar s_m \gamma^\mu_L s_m}{2}
+ \bar d_m \gamma^\mu_L s_m + \bar s_m \gamma^\mu_L d_m \right)\right],
\end{eqnarray}
in which all components of the triplet of $SU(2)$ currents $\left[
\frac12\left(\bar d_m \gamma^\mu_L d_m - \bar s_m \gamma^\mu_L s_m\right),
\bar d_m \gamma^\mu_L s_m, \bar s_m \gamma^\mu_L d_m\right]$, the
corresponding (vector) charges of which make up the relevant $SU(2)$ algebra,
have the same coefficient $(\alpha - \beta)$.
The work \cite{DuretMachet2} states accordingly that 
the ``Cabibbo angle'' is controlled by this $SU(2)$ symmetry.

The generalization to three generations is now straightforward.
Neutral currents  are controlled by the
product $K^\dagger K$ of the $3 \times 3$ mixing matrix $K$ with its
hermitian conjugate; for example, the (left-handed) neutral
currents for quarks with electric charge $(-1/3)$ read
\begin{equation}
\overline{\left(\begin{array}{c} d_f \cr s_f \cr b_f \end{array}\right)}
\gamma^\mu_L\left(\begin{array}{c} d_f \cr s_f \cr b_f \end{array}\right)
=
\overline{\left(\begin{array}{c} d_m \cr s_m \cr b_m \end{array}\right)}
\gamma^\mu_L\; K^\dagger_d K_d
\left(\begin{array}{c} d_m \cr s_m \cr b_m \end{array}\right).
\end{equation}
Requesting $SU(2)$ symmetry in each $(i,j)$ channel is trivially
equivalent to the condition that, in this channel, universality for the
diagonal currents is violated with the same strength as the absence of
non-diagonal currents. We will show that all presently known mixing angles,
in the quark as well as in the leptonic sectors, satisfy this criterion.
  
\subsection{Mixing matrices. Notations}

We write each mixing matrix  $K$ as a product of three 
matrices, which reduce,
in the unitarity limit, to the basic rotations by $\theta_{12}$,
$\theta_{23}$ and $\theta_{13}$ (we are not concerned with $CP$ violation)
\begin{equation}
K = \left(\begin{array}{ccc} 1 & 0 & 0 \cr
                             0 & c_{23} & s_{23} \cr
                             0 & - \tilde{s}_{23} & \tilde{c}_{23}
\end{array}\right) \times 
\left(\begin{array}{ccc} c_{13} & 0 & s_{13} \cr
                         0 & 1 & 0 \cr
                         - \tilde{s}_{13}& 0 & \tilde{c}_{13}
\end{array}\right) \times
\left(\begin{array}{ccc} c_{12} & s_{12} & 0 \cr
                         - \tilde{s}_{12} & \tilde{c}_{12} & 0 \cr
                         0 & 0 & 1
\end{array}\right).
\end{equation}
We parametrize each basic matrix, which is {\em a priori} non-unitary, 
with two angles, respectively
 $(\theta_{12}, \tilde{\theta}_{12})$, $(\theta_{23},
\tilde{\theta}_{23})$ and $(\theta_{13}, \tilde{\theta}_{13})$.
We deal accordingly with six mixing
angles, instead of three in the unitary case (where
$\tilde{\theta}_{ij} = \theta_{ij}$). 
We will use throughout the paper the notations 
$s_{ij}= \sin(\theta_{ij}), \tilde{s}_{ij} = \sin(\tilde{\theta}_{ij})$,
and likewise, for the cosines, 
$c_{ij} = \cos(\theta_{ij}), \tilde{c}_{ij} = \cos(\tilde{\theta}_{ij})$. 

To lighten the text, the elements of $K^\dagger K$ will be abbreviated
 by $[ij], i,j=1\ldots 3$ instead of $(K^\dagger K)_{[ij]}$, and the
corresponding neutral current will be noted $\{ij\}$. So, in the quark case,
$\{12\}$ stands for $\bar u_m \gamma^\mu_L c_m$ or
$ \bar d_m \gamma^\mu_L s_m$, and, in
the neutrino case, for $\bar\nu_{em} \gamma^\mu_L \nu_{\mu m}$ or $\bar
e_m \gamma^\mu_L \mu_m$.

\subsection{The unitarization conditions}

They are five: three arise from the
absence of non-diagonal neutral currents for mass eigenstates, and two from
the universality of diagonal currents.  Accordingly,
one degree of freedom is expected to be unconstrained.

\subsubsection{Absence of non-diagonal neutral currents of mass eigenstates}

The three conditions read:\newline

\vbox{
$\ast$ for the absence of $\{13\}$ and $\{31\}$ currents:
\begin{equation}
[13]=0=[31] \Leftrightarrow
c_{12}\left[c_{13}s_{13} -\tilde c_{13} \tilde s_{13}(\tilde c_{23}^2 + s_{23}^2)\right] 
- \tilde c_{13} \tilde s_{12}(c_{23} s_{23} - \tilde c_{23} \tilde s_{23}) = 0;
\label{eq:nodb}
\end{equation}
$\ast$ for the absence of $\{23\}$ and $\{32\}$ currents:
\begin{equation}
[23]=0=[32] \Leftrightarrow
s_{12}\left[c_{13}s_{13} -\tilde c_{13} \tilde s_{13}(\tilde c_{23}^2 + s_{23}^2)\right] 
+ \tilde c_{13} \tilde c_{12}(c_{23} s_{23} - \tilde c_{23} \tilde s_{23}) = 0;
\label{eq:nosb}
\end{equation}
$\ast$ for the absence of $\{12\}$ and $\{21\}$ currents:
\begin{eqnarray}
&& [12]=0=[21] \Leftrightarrow\cr
&&s_{12}c_{12} c_{13}^2 - \tilde s_{12} \tilde c_{12}(c_{23}^2 + \tilde s_{23}^2) + s_{12} c_{12} \tilde
s_{13}^2 ( s_{23}^2 + \tilde c_{23}^2) + \tilde s_{13} (s_{12} \tilde s_{12} - c_{12} \tilde
c_{12})(c_{23} s_{23} - \tilde c_{23} \tilde s_{23})=0.\cr
&&
\label{eq:nods}
\end{eqnarray}
}

\subsubsection{Universality of diagonal neutral currents of mass eigenstates}

The two independent conditions read:

$\ast$ equality of $\{11\}$ and $\{22\}$ currents:
\begin{eqnarray}
&&[11]-[22]=0 \Leftrightarrow \cr
&&(c_{12}^2 - s_{12}^2)\left[ c_{13}^2 + \tilde s_{13}^2(s_{23}^2 + \tilde c_{23}^2)\right]
-(\tilde c_{12}^2 - \tilde s_{12}^2)(c_{23}^2+ \tilde s_{23}^2)\cr
&& \hskip 3cm + 2\tilde s_{13}(c_{23}s_{23}
- \tilde c_{23} \tilde s_{23})(c_{12} \tilde s_{12} + s_{12} \tilde
c_{12})=0;
\label{eq:ddss}
\end{eqnarray}

$\ast$ equality of $\{22\}$ and $\{33\}$ currents:
\begin{eqnarray}
&&[22]-[33]=0 \Leftrightarrow \cr
&&s_{12}^2 + \tilde c_{12}^2(c_{23}^2 + \tilde s_{23}^2) - (s_{23}^2 + \tilde c_{23}^2)
+(1+s_{12}^2)\left[ \tilde s_{13}^2(s_{23}^2 + \tilde c_{23}^2) -s_{13}^2
\right]\cr
&&\hskip 6cm
+ 2s_{12} \tilde s_{13} \tilde c_{12}(\tilde c_{23} \tilde s_{23} - c_{23} s_{23}) =0.
\label{eq:ssbb}
\end{eqnarray}
The equality  of $\{11\}$ and $\{33\}$ currents is of course
 not an independent condition.

\subsection{Solutions for $\boldsymbol{\theta_{13} = 0 =
\tilde{\theta}_{13}}$}
\label{subsection:vanish}

In a first step, to ease solving the system of trigonometric equations,
we shall study the configuration in which one of the two angles
parametrizing the 1-3 mixing vanishes  
\footnote{By doing so, we exploit the possibility to fix one degree of
freedom left {\em a priori} unconstrained by the five equations; see
subsection \ref{subsection:NCMM}.},
 which is very close to what is
observed experimentally in the quark sector, and likely in the neutrino
sector. It turns out, as demonstrated in Appendix \ref{section:theta3},
that the second mixing angle vanishes simultaneously.
We accordingly work in the approximation (the sensitivity of the
solutions to a small variation of $\theta_{13}, \tilde{\theta}_{13}$
will be studied afterwards)
\begin{equation}
\theta_{13} = 0 =\tilde{\theta}_{13}.
\label{eq:cond1}
\end{equation}
Eqs. (\ref{eq:nodb}), (\ref{eq:nosb}), (\ref{eq:nods}), (\ref{eq:ddss}) and
(\ref{eq:ssbb}),  reduce in this limit to

\vbox{
\begin{subequations}\label{subeq:geneq0}
\begin{equation}
 - \tilde s_{12}(c_{23} s_{23} - \tilde c_{23} \tilde s_{23}) = 0,
\label{eq:nodb0}
\end{equation}
\begin{equation}
 \tilde c_{12}(c_{23} s_{23} - \tilde c_{23} \tilde s_{23}) = 0,
\label{eq:nosb0}
\end{equation}
\begin{equation}
s_{12} c_{12}  - \tilde s_{12} \tilde c_{12}(c_{23}^2 + \tilde s_{23}^2) = 0,
\label{eq:nods0}
\end{equation}
\begin{equation}
(c_{12}^2 -s_{12}^2) -(\tilde c_{12}^2 - \tilde s_{12}^2)(c_{23}^2 + \tilde s_{23}^2) = 0,
\label{eq:ddss0}
\end{equation}
\begin{equation}
s_{12}^2 + \tilde c_{12}^2(c_{23}^2 + \tilde s_{23}^2) - (s_{23}^2 + \tilde c_{23}^2)  = 0.
\label{eq:ssbb0}
 \end{equation}
\end{subequations}
}

It is shown in Appendix \ref{section:sol0} that the only solutions are
$\theta_{12}$ and $\theta_{23}$ 
Cabibbo-like ($\tilde{\theta}_{12,23} = \theta_{12,23} + k\pi$) or maximal
($\theta_{12,23} = \pi/4 + n\pi/2,\  \tilde{\theta}_{12,23}
= \pi/4 + m\pi/2$).

Accordingly, the two following sections will respectively start from:

$\ast$ $\theta_{12}$ and $\theta_{23}$ Cabibbo-like (and, in a first step,
 vanishing
$\theta_{13}$), which finally leads to a mixing pattern similar to what is
observed for quarks;

$\ast$  $\theta_{23}$ maximal and $\theta_{12}$ Cabibbo like (and, in a
first step, vanishing
$\theta_{13}$), which finally leads to a mixing pattern similar to the one
observed for neutrinos.

\section{The quark sector; constraining the three CKM angles}

Mass splittings entail that the previous general conditions, which, when
exactly satisfied, correspond {\em de facto}
to unitary mixing matrices, cannot be exactly fulfilled.
We  investigate the vicinity of their solutions, and show that the same
violation pattern that led to an accurate determination of the
Cabibbo angle
in the case of two generations, is also satisfied by the
CKM angles in the case of three generations.

\subsection{The simplified case $\boldsymbol{\theta_{13} = 0 =\tilde{\theta}_{13}}$}
\label{subsec:qapp}

In the neighborhood of the  solution with both $\theta_{12}$ and
$\theta_{23}$ Cabibbo-like, we write
\begin{eqnarray}
\tilde{\theta}_{12} &=& \theta_{12} + \epsilon,\cr
\tilde{\theta}_{23} &=& \theta_{23} + \eta.
\label{eq:q0}
\end{eqnarray}
The pattern $(\theta_{13} = 0 = \tilde{\theta}_{13})$ can be reasonably
considered to be  close to the experimental situation,
at least close enough for
trusting not only the relations involving the first and second generation,
but also the third one.

Like in \cite{DuretMachet2}, we impose that
the absence of $\{12\}, \{21\}$ neutral currents is violated with the
same strength as the universality of $\{11\}$ and $\{22\}$ currents.
It reads
\begin{equation}
|2\eta s_{12} c_{12} s_{23} c_{23} + \epsilon (c_{12}^2 -s_{12}^2)|
= |-2\eta s_{23} c_{23}(c_{12}^2 -s_{12}^2) + 4\epsilon s_{12} c_{12}|.
\label{eq:q1}
\end{equation}
We choose the ``$+$'' sign for both sides, such that, for two generations only,
the Cabibbo angle satisfies $\tan(2\theta_{12}) = + 1/2$. (\ref{eq:q1}) yields
the ratio $\eta/\epsilon$, that we then plug into the condition
 equivalent to (\ref{eq:q1}) for the $(2,3)$ channel.
\begin{equation}
|\eta c_{12}(c_{23}^2 - s_{23}^2)| = | 2\eta s_{23}c_{23}(1+c_{12}^2) 
-2\epsilon s_{12} c_{12}|.
\label{eq:q2}
\end{equation}
(\ref{eq:q1}) and (\ref{eq:q2}) yield
\begin{equation}
\tan(2\theta_{23}) = \displaystyle\frac{c_{12}}{1+c_{12}^2 -
2s_{12}c_{12}\displaystyle\frac{(s_{12}c_{12} + c_{12}^2 -
s_{12}^2)}{4s_{12}c_{12} -(c_{12}^2 -s_{12}^2)}}
\approx \displaystyle\frac{c_{12}}{2 -\displaystyle
\frac54\displaystyle\frac{s_{12}c_{12}}{\tan(2\theta_{12})
-\displaystyle\frac12}}.
\label{eq:q3}
\end{equation}
In the r.h.s. of (\ref{eq:q3}), we have assumed that $\theta_{12}$ is close to
its Cabibbo value $\tan(2\theta_{12}) \approx 1/2$. $\theta_{23}$ is  seen
to vanish with $[\tan(2\theta_{23}) -1/2]$.
The predicted value for $\theta_{23}$ is plotted in Fig.~1  as a
function of $\theta_{12}$, together with the experimental intervals for
$\theta_{23}$ and $\theta_{12}$. There are two \cite{PDG}
 for $\theta_{12}$; the first comes
from the measures of $V_{ud}$ (in black on Fig.~1)
\begin{equation}
V_{ud} \in [0.9735,0.9740] \Rightarrow \theta_{12} \in [0.2285,0.2307],
\label{eq:Vud}
\end{equation}
and the second from the measures of $V_{us}$ (in purple on Fig.~1)
\begin{equation}
V_{us} \in [0.2236,0.2278] \Rightarrow \theta_{12} \in [0.2255, 0.2298].
\label{eq:Vus}
\end{equation}
\begin{center}
\includegraphics[height=7truecm,width=9truecm]{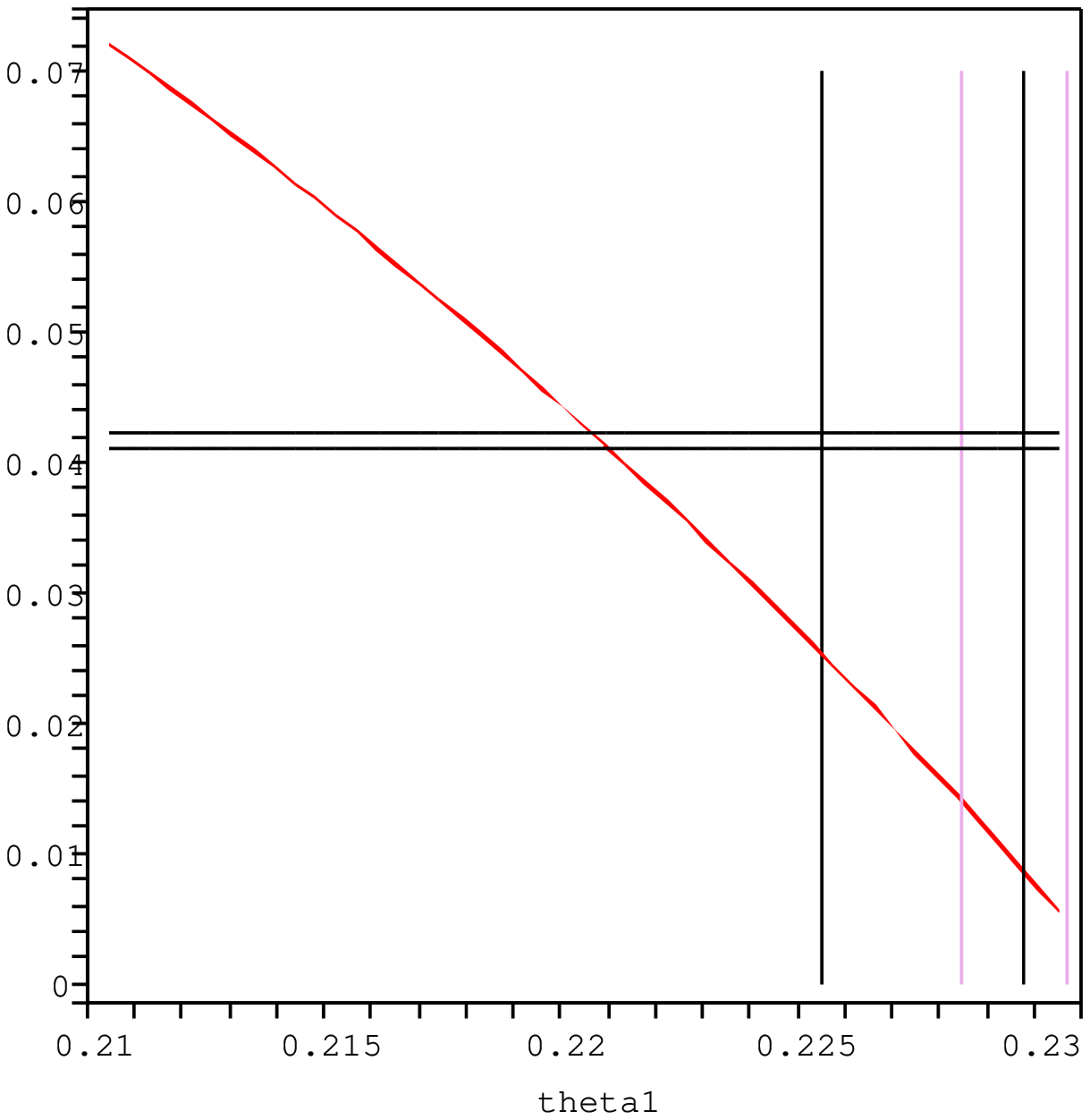}
\figskip
{\em Fig.~1: $\theta_{23}$ for quarks as a function of $\theta_{12}$;
simplified case $\theta_{13}=0=\tilde\theta_{13}$ }
\end{center}
\figskip

The measured value for $\theta_{23}$ is seen on Fig.~1
 to correspond to $\theta_{12}
\approx 0.221$, that is $\cos (\theta_{12}) \approx 0.9757$.
Our prediction for $\cos(\theta_{12})$ is accordingly $1.7\,10^{-3}$
 away from
the upper limit of the present upper bound for  $V_{ud} \equiv c_{12}c_{13}$
\cite{Vud} \cite{PDG}; it
corresponds to twice the experimental uncertainty.
It also  corresponds to $\sin(\theta_{12}) = 0.2192$,
while $V_{us} \equiv s_{12} c_{13}$ is measured to be $0.2247(19)$
 \cite{KLOE} \cite{PDG};
there, the discrepancy is $2/100$, only slightly above the $1.8/100$
relative width of the experimental interval.

The approximation which sets $\theta_{13} = 0 =\tilde{\theta}_{13}$ is accordingly
reasonable,
though it yields results slightly away from experimental bounds.
We show in the next subsection that relaxing this approximation gives
results in excellent agreement with present experiments.

\subsection{Going to $\boldsymbol{(\theta_{13}\not=0,
\tilde{\theta}_{13}\not=0)}$}

Considering all angles to be Cabibbo-like with, in addition to (\ref{eq:q0})

\vbox{
\begin{equation}
\tilde{\theta}_{13} = \theta_{13} + \rho,
\label{eq:q4}
\end{equation}
the l.h.s.'s of eqs. (\ref{eq:nodb}),(\ref{eq:nosb}),(\ref{eq:nods}),
(\ref{eq:ddss}), (\ref{eq:ssbb}) and the sum (\ref{eq:ddss} +
\ref{eq:ssbb}) depart respectively from zero by
\begin{subequations}
\label{subeq:general}
\begin{equation}
\eta c_{13} \left[s_{12}(c_{23}^2 - s_{23}^2) +
         2s_{13} c_{12} c_{23} s_{23}\right]
- \rho c_{12} (c_{13}^2 - s_{13}^2);
\label{eq:nodb3}
\end{equation}
\begin{equation}
\eta c_{13}\left[ - c_{12} (c_{23}^2 - s_{23}^2)
        + 2 s_{13} s_{12} c_{23} s_{23} \right]
- \rho s_{12} (c_{13}^2 - s_{13}^2);
\label{eq:nosb3}
\end{equation}
\begin{equation}
-\epsilon(c_{12}^2-s_{12}^2) + \eta\left[
s_{13}(c_{23}^2- s_{23}^2)(c_{12}^2
- s_{12}^2) -2 c_{23} s_{23} c_{12} s_{12} (1 + s_{13}^2) \right]
+ 2\rho c_{13} s_{13} c_{12} s_{12};
\label{eq:nods3}
\end{equation}
\begin{equation}
4\epsilon c_{12}s_{12}
+\eta \left[-4 s_{13} s_{12}c_{12}(c_{23}^2- s_{23}^2)
-2 c_{23} s_{23} (c_{12}^2 - s_{12}^2)(1 + s_{13}^2) \right]
+ 2\rho c_{13} s_{13} (c_{12}^2 - s_{12}^2 );
\label{eq:ddss3}
\end{equation}
\begin{equation}
-2\epsilon s_{12}c_{12}
+ \eta\left[ 2s_{13} c_{12} s_{12} (c_{23}^2 - s_{23}^2) + 2c_{23}s_{23}
\left((c_{12}^2 - s_{12}^2) + c_{13}^2 ( 1+s_{12}^2) \right)\right]
+ 2\rho c_{13} s_{13} (1+s_{12}^2); 
\label{eq:ssbb3}
\end{equation}
\begin{equation}
2\epsilon s_{12}c_{12}
+ \eta\left[ -2 s_{13} c_{12} s_{12} (c_{23}^2 - s_{23}^2) 
+ 2 c_{23} s_{23}
\left( c_{13}^2 (1+c_{12}^2) -(c_{12}^2 - s_{12}^2 ) \right) \right]
+ 2 \rho c_{13} s_{13} (1+ c_{12}^2).
\label{eq:ddbb3}
\end{equation}
\end{subequations}
}

We have added (\ref{eq:ddbb3}), which is not an independent relation, but
the sum of (\ref{eq:ddss3}) and (\ref{eq:ssbb3}); it expresses the
violation in the universality of diagonal $\{11\}$ and $\{33\}$
currents.

\subsubsection{A guiding calculation}
\label{subsub:guide}

Before doing the calculation in full generality, and to make a clearer
difference with the neutrino case, we first do it in the limit where one
neglects terms which are
quadratic in the small quantities $\theta_{13}$ and $\rho$.
By providing simple intermediate formul\ae, it enables in particular to
suitably choose the signs which occur in equating the moduli of two
quantities.  Eqs.(\ref{subeq:general}) become
\begin{subequations}
\label{subeq:simple}
\begin{equation}
\eta \left[s_{12}(c_{23}^2 - s_{23}^2)
+ 2s_{13} c_{12} c_{23} s_{23}\right] - \rho c_{12};
\label{eq:nodb3s}
\end{equation}
\begin{equation}
\eta \left[ - c_{12} (c_{23}^2 - s_{23}^2)
+ 2 s_{13} s_{12} c_{23} s_{23} \right] - \rho s_{12};
\label{eq:nosb3s}
\end{equation}
\begin{equation}
-\epsilon(c_{12}^2-s_{12}^2)
+ \eta\left[ s_{13}(c_{23}^2- s_{23}^2)(c_{12}^2 - s_{12}^2)
-2 c_{23} s_{23} c_{12} s_{12} \right];
\label{eq:nods3s}
\end{equation}
\begin{equation}
4\epsilon c_{12}s_{12} - 2\eta \left[2 s_{13} s_{12}c_{12}(c_{23}^2
- s_{23}^2) + c_{23} s_{23} (c_{12}^2 - s_{12}^2) \right];
\label{eq:ddss3s}
\end{equation}
\begin{equation}
-2\epsilon s_{12}c_{12} + 2\eta\left[ s_{13} c_{12} s_{12} (c_{23}^2
- s_{23}^2) + c_{23}s_{23} (1 +c_{12}^2 )\right]; 
\label{eq:ssbb3s}
\end{equation}
\begin{equation}
2\epsilon s_{12}c_{12} + 2\eta\left[ - s_{13} c_{12} s_{12} (c_{23}^2 
- s_{23}^2)  +  c_{23} s_{23} ( 1+ s_{12}^2)  \right].
\label{eq:ddbb3s}
\end{equation}
\end{subequations}
The principle of the method is the same as before.
From (\ref{eq:nods3s}) = (-)(\ref{eq:ddss3s})
\footnote{The (-) signs ensures that $\tan(2\theta_{12}) \approx (+) 1/2$.}
, which
expresses that the absence of non-diagonal $\{12\}$ current is violated
with the same strength as the universality of $\{11\}$ and $\{22\}$
currents, one gets $\epsilon/\eta$ as a
function of $\theta_{12}, \theta_{23}, \theta_{13}$
\footnote{
\begin{equation}
\frac{\epsilon}{\eta} = s_{13}(c_{23}^2-s_{23}^2) + 2s_{23}c_{23}
\frac{s_{12}c_{12}+c_{12}^2-s_{12}^2}{4c_{12}s_{12} -(c_{12}^2-s_{12}^2)};
\label{eq:epsq}
\end{equation}
$\epsilon /\eta$ has a pole at $\tan(2\theta_{12}) = 1/2$, the predicted value
of the Cabibbo angle for two generations.
}.
 This expression is plugged in the relation 
(\ref{eq:nosb3s}) = (-)(\ref{eq:ssbb3s})\footnote{There, again, the (-)
sign has to be chosen so as to recover approximately (\ref{eq:q3}).},
which expresses the same condition for the $(2,3)$ channel;
from this, one 
extracts $\rho/\eta$ as a function of $\theta_{12}, \theta_{23},
\theta_{13}$
\footnote{
\begin{equation}
\displaystyle\frac{\rho}{\eta} = 2  c_{23} s_{23}\left[ s_{13}
-c_{12}
\left( 2\displaystyle\frac
{(c_{12}s_{12}+c_{12}^2-s_{12}^2)}{4s_{12}c_{12} -(c_{12}^2-s_{12}^2)} -
\displaystyle\frac{1+c_{12}^2}{c_{12}s_{12}}
+\displaystyle\frac{1}{s_{12}}
\displaystyle\frac{c_{23}^2-s_{23}^2}{2s_{23}c_{23}}\right)\right].
\label{eq:rhoq}
\end{equation}
$\rho/\eta$ has a pole at $\tan(2\theta_{12}) = 1/2$ 
and, for $\theta_{13}=0$, it vanishes, as expected, when
$\theta_{12}$ and $\theta_{23}$ satisfy the relation (\ref{eq:q3}),
which has been deduced for $\tilde{\theta}_{13}
(\equiv \theta_{13}+\rho)=0=\theta_{13}$.}.
The expressions that have been obtained
 for $\epsilon/\eta$ and $\rho/\eta$ are then
inserted into the third relation, \vline\ (\ref{eq:nodb3s})\ \vline\ =
\vline\ (\ref{eq:ddbb3s})\ \vline\ , which now corresponds to the $(1,3)$
channel.  This last step
yields a relation $F_0(\theta_{12},\theta_{23},\theta_{13})=1$
 between the three angles $\theta_{12},\theta_{23}, \theta_{13}$.

It turns out that $\frac{\partial F_0(\theta_{12},\theta_{23},\theta_{13})}{\partial
\theta_{13}}=0$, such that, in this case, a  condition between
  $\theta_{12}$ and $\theta_{23}$ alone eventually fulfills the three relations
under concern
\begin{equation}
1= \left|\frac{\text{viol}([11] = [22])}{\text{viol}([12] =0 = [21])}\right|
=\left|\frac{\text{viol}([22] = [33])}{\text{viol}([23] =0=[32])}\right|
=\left|\frac{\text{viol}([11] = [33])}{\text{viol}([13] =0=[31])}\right|
\Leftrightarrow \tilde F_0(\theta_{12},\theta_{23}) =1.
\label{eq:cond0}
\end{equation}
\vbox{
\begin{center}
\includegraphics[height=7truecm,width=9truecm]{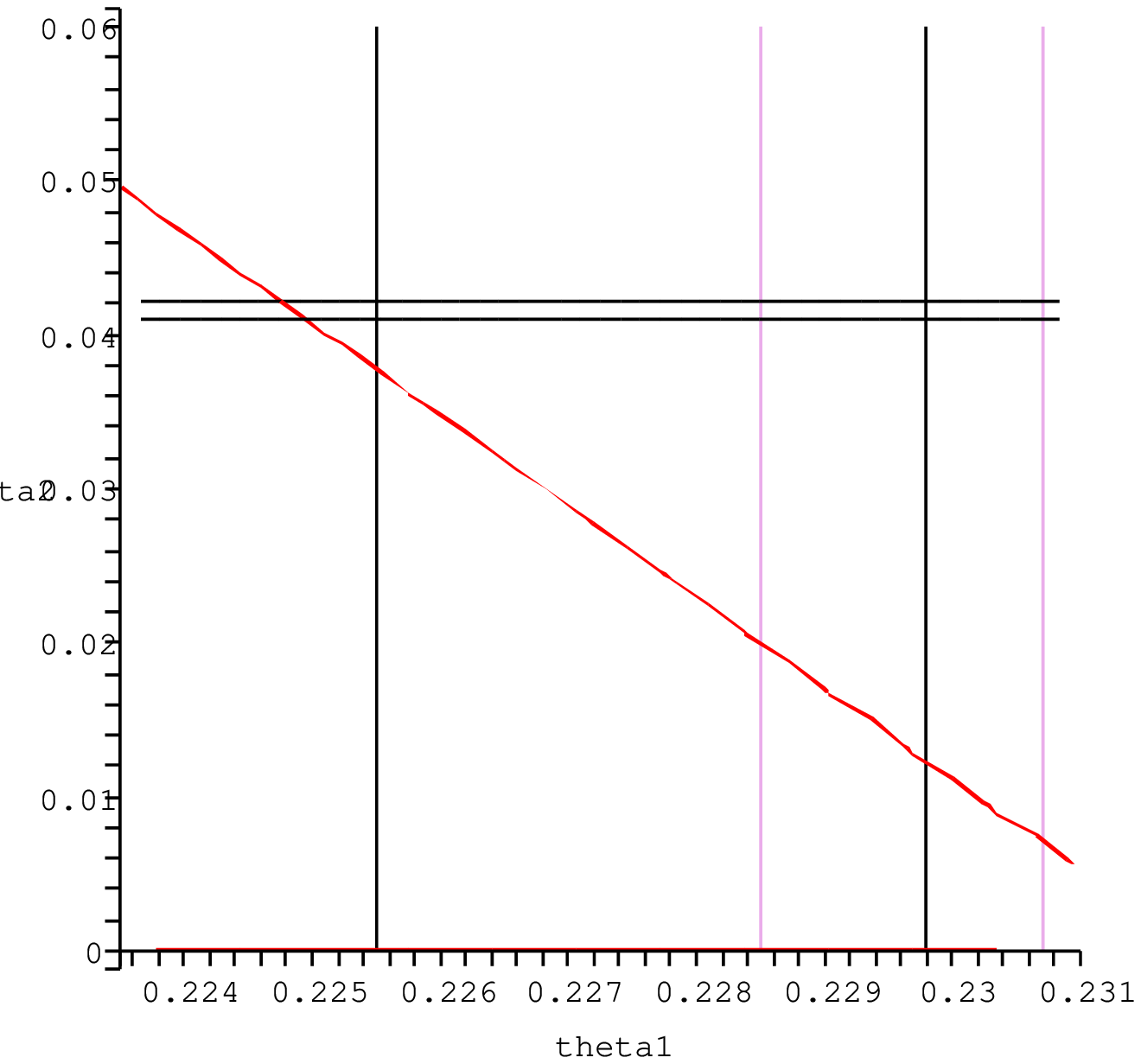}

{\em Fig.~2: $\theta_{23}$ for quarks
 as a function of $\theta_{12}$; neglecting terms
quadratic in $\theta_{13}$}
\end{center}
}
\figskip

$\theta_{23}$ is plotted on Fig.~2 as a function of $\theta_{12}$, together with the
experimental intervals for $\theta_{23}$ and $\theta_{12}$ (the intervals
for $\theta_{12}$  come respectively from $V_{ud}$ (eq.~(\ref{eq:Vud}))
and $V_{us}$ (eq.~(\ref{eq:Vus}))).

The precision obtained is much better than  in Fig.~1 since, in
particular, for $\theta_{23}$ within its experimental range,
the discrepancy
between the predicted $\theta_{12}$ and its lower experimental limit
coming from $V_{us}$ is smaller than the two experimental intervals, and
even smaller than their intersection.

\subsubsection{The general solution}
\label{subsub:general}

The principle for solving the general equations (\ref{subeq:general})
is the same as above.
One first uses the relation  (\ref{eq:nods3}) = (-) (\ref{eq:ddss3})
 to determine $\rho/\epsilon$ in terms of $\eta/\epsilon$.
The result is plugged in the relation (\ref{eq:nosb3}) = (-)
(\ref{eq:ssbb3}), which fixes $\eta/\epsilon$, and thus
$\rho/\epsilon$ as  functions of $(\theta_{12},\theta_{23},\theta_{13})$. These
expressions for $\eta/\epsilon$ and $\rho/\epsilon$ are finally plugged in
the relation \vline\  (\ref{eq:nodb3})\ \vline\ = \vline\ 
(\ref{eq:ddbb3})\ \vline\  , which provides a condition
$F(\theta_{12},\theta_{23},\theta_{13})=1$.
When it is fulfilled, the universality of
each pair of diagonal neutral currents of mass eigenstates
and the absence of the corresponding non-diagonal currents are violated
with the same strength, in the three channels
$(1,2)$, $(2,3)$ and $(1,3)$. 

The results are displayed in Fig.~3; $\theta_{23}$ is plotted
as a function of $\theta_{12}$ for $\theta_{13} = 0.004$ and $0.01$.
The present experimental interval is \cite{PDG}
\begin{equation}
V_{ub} = \sin(\theta_{13}) \approx \theta_{13} \in [4\,10^{-3},4.6\,10^{-3}].
\label{eq:theta3}
\end{equation}
\vbox{
\begin{center}
\includegraphics[height=7truecm,width=9truecm]{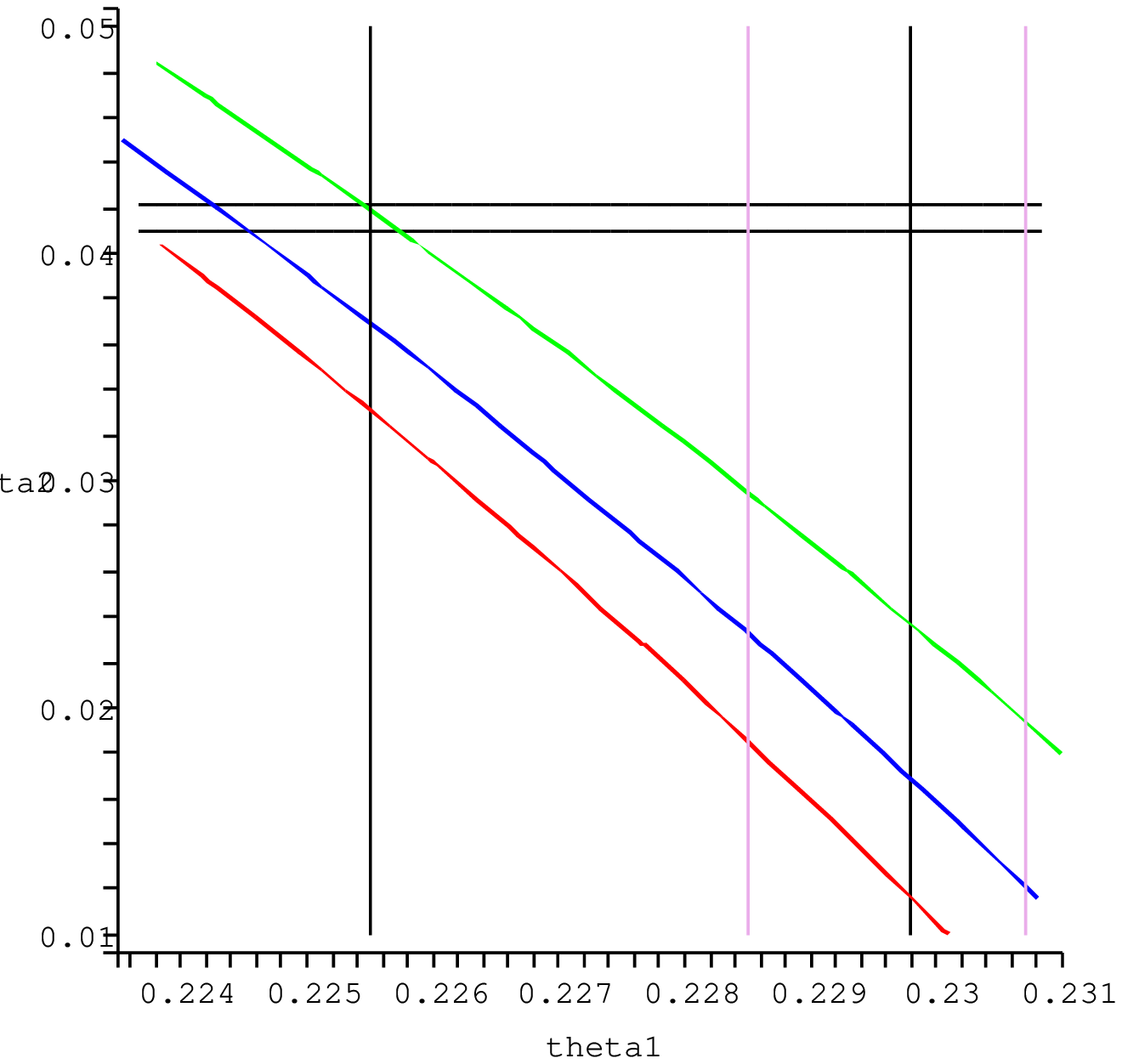}

{\em Fig.~3: $\theta_{23}$ for quarks as a function of $\theta_{12}$,
general case. 
$\theta_{13}=0$ (red), $0.004$ (blue) and $0.01$ (green)}
\end{center}}
\figskip
We conclude that:

$\ast$ The discrepancy between our predictions and experiments is
smaller than the experimental uncertainty;

$\ast$ a slightly larger value of $\theta_{13}$ and/or
 slightly smaller values of $\theta_{23}$ and/or $\theta_{12}$
still increase the agreement between our
predictions and experimental measurements;

$\ast$ the determination of $\theta_{12}$ from $V_{us}$ seems
preferred to that  from $V_{ud}$.

Another  confirmation of the relevance of our criterion
 is given in the next section concerning neutrino mixing angles.

\section{A neutrino-like pattern; quark-lepton complementarity}
\label{section:neutrinos}

In the ``quark case'', we dealt with three ``Cabibbo-like'' angles.
The configuration that we investigate here is the one  in which
$\theta_{23}$ is, as observed experimentally \cite{PDG}, (close to)
 maximal, and $\theta_{12}$ and $\theta_{13}$ are
Cabibbo-like (see subsection \ref{subsection:vanish}). 

\subsection{The  case $\boldsymbol{\theta_{13} = 0 = \tilde{\theta}_{13}}$}
\label{subsec:nu1}

We explore the vicinity of this solution,
slightly departing from the corresponding unitary mixing matrix,
by considering that $\tilde{\theta}_{12}$ now slightly differs
from $\theta_{12}$, and $\tilde{\theta}_{23}$
from its maximal value
\begin{eqnarray}
\tilde{\theta}_{12} &=& \theta_{12} + \epsilon,\cr
\theta_{23} = \pi/4 &,&  \tilde{\theta}_{23} = \theta_{23} + \eta.
\label{eq:eta}
\end{eqnarray}
The l.h.s.'s of eqs. (\ref{eq:nodb}) (\ref{eq:nosb}) (\ref{eq:nods})
(\ref{eq:ddss}) and (\ref{eq:ssbb}) no longer vanish, and become
respectively 

\vbox{
\begin{subequations}
\label{subeq:nueqs}
\begin{equation}
-\frac 12 \eta^2 (s_{12} + \epsilon c_{12}),
\label{eq:neut1}
\end{equation}
\begin{equation}
\frac 12 \eta^2 (c_{12} - \epsilon s_{12}),
\label{eq:neut2}
\end{equation}
\begin{equation}
\ast\ -\eta s_{12}c_{12} + \epsilon (s_{12}^2 -c_{12}^2)(1+\eta),
\label{eq:neut3}
\end{equation}
\begin{equation}
\ast\ -\eta (c_{12}^2 -s_{12}^2) + 4\epsilon s_{12}c_{12}(1+\eta),
\label{eq:neut4}
\end{equation}
\begin{equation}
\eta (1+c_{12}^2) -2\epsilon s_{12}c_{12}(1+\eta),
\label{eq:neut5}
\end{equation}
\end{subequations}
}

showing by which amount the five conditions under scrutiny are now 
violated. Some care has to be taken concerning the accurateness of
equations (\ref{subeq:nueqs}).
Indeed, we imposed a value of $\theta_{13}$ which is 
probably not the physical one (even if close to). It is then reasonable to
consider that channel $(1,2)$ is the less sensitive to this approximation
and that, accordingly,
 of the five equations above, (\ref{eq:neut3}) and (\ref{eq:neut4}),
marked with an ``$\ast$'', are  the most accurate
\footnote{The limitation of this approximation also appears in the fact
that  (\ref{eq:neut2}), of second order in $\eta$, is
not compatible with (\ref{eq:neut5}), which is of first order.}
.

The question: is there a special value of $\theta_{12} = \tilde{\theta}_{12}$
Cabibbo-like for which  small deviations $(\epsilon, \eta)$ from 
unitarity  entail equal strength violations of \newline
$\ast$ the absence of $\{12\}, \{21\}$ non-diagonal neutral
currents;\newline
$\ast$ the universality of $\{11\}$ and $\{22\}$ neutral currents ?

gets then a simple answer
\begin{equation}
s_{12}c_{12} = c_{12}^2 - s_{12}^2 \Rightarrow \tan(2\theta_{12}) = 2.
\end{equation}
We did not take into account the terms proportional to 
$\epsilon$ because we assumed that the mass splittings
between the first and second generations
(from which the lack of unitarity originates)
are much smaller that the ones  between
the second and the third generation
\footnote{Since the three angles play {\em a priori} symmetric roles,
the simultaneous vanishing of $\theta$ and $\tilde{\theta}$, which we
demonstrated for $\theta_{13}$ and $\tilde{\theta}_{13}$ (see Appendix
\ref{section:theta3}), should also occur for the other
angles. Two competing effects accordingly contribute to the magnitude of
the parameters $\epsilon$, $\eta$ \ldots: on one hand,
they should be proportional to
(some power of) the corresponding $\theta$, and, on the other hand, one
reasonably expects them to increase with the mass
splitting between the fermions mixed by this $\theta$. So, in the quark
sector, that the violation
of unitarity should be maximal for $\theta_{13}$ is not guaranteed since
the corresponding mixing angle is also very small (as expected from
hierarchical mixing matrices \cite{MachetPetcov}). A detailed
investigation of this phenomenon is postponed to a further work.
In the neutrino sector, however, since $\theta_{23}$ is maximal (large), the
assumption that the mass splitting between the second and third generation
is larger than between the first and second is enough to guarantee 
$\epsilon \ll \eta$.}.

In the case of two generations, only $\epsilon$ appears, and
one immediately recovers from (\ref{eq:neut3}) and (\ref{eq:neut4}) the
condition fixing $\tan(2\theta_c) = 1/2$ for the Cabibbo angle.

Accordingly, the same type of requirement that led to a value of the
Cabibbo angle for two generations very close to the observed value leads,
for three generations, to a value of the first mixing angle satisfying
the quark-lepton complementarity relation (\ref{eq:qlc}) \cite{QLC}.

The values of $\theta_{12}$ and $\theta_{23}$  determined through this
procedure are very close to
the observed neutrino mixing angles \cite{PDG}. 

Though we only considered the two equations that are {\em a priori}
the least sensitive to our choice of a vanishing third mixing angle
(which is not yet confirmed experimentally),
it is instructive to investigate the sensitivity of
our solution to a small non-vanishing value of $\theta_{13}$.
This is done in Appendix \ref{section:maxsens} in which, for this purpose,
we  made the simplification $\tilde{\theta}_{13} \approx \theta_{13}$.
It turns out
that the terms proportional to $s_{13}$ in the two equations
$[12]=0=[21]$ and $\vline\ [11]\ \vline =\ \vline[22]\ \vline$ are also
proportional to $(c_{23}^2 - s_{23}^2)$, such that our solution
with $\theta_{23}$ maximal is very stable with
respect to a variation of $\theta_{13}$ around zero.
This may of course not be the case
for the other three equations, which are expected to be more sensitive to
the value of $\theta_{13}$.

\subsection{Prediction for $\boldsymbol{\theta_{13}}$}
\label{subsec:nugen}

We now consider, like we did for quarks,
the general case $\theta_{13} \not= 0
\not = \tilde{\theta}_{13} (\rho\not=0)$, $\tilde{\theta}_{12} \not= \theta_{12} (\epsilon \not=
0)$, $\tilde{\theta}_{23} \not= \theta_{23} (\eta\not=0)$,  while
assigning to $\theta_{12}$ and $\theta_{23}$
their values predicted in subsection \ref{subsec:nu1}.

We investigate the eight different relations between
$\theta_{12}$, $\theta_{23}$
and $\theta_{13}$ which originate from the $2 \times 2 \times 2$ possible
 sign combinations in the conditions (\ref{eq:cond0}) (the r.h.s. is now
replaced by a condition $F(\theta_{12}, \theta_{23}, \theta_{13})=1$
involving the three mixing angles),
where each  modulus can be alternatively replaced by ``$+$'' or ``$-$''.

Among the solutions found for $\theta_{13}$,  only two (up to a
sign) satisfy the very loose experimental bound
\begin{equation}
\sin^2 (\theta_{13}) \leq 0.1.
\end{equation}
They correspond respectively to the sign combinations
$(+/-/-)$, $(+/+ /+)$, $(-/+/+)$ and $(-/-/-)$
\begin{eqnarray}
\theta_{13} = \pm 0.2717 &,& \sin^2(\theta_{13}) = 0.072,\cr
&&\cr
\theta_{13} = \pm 5.7\,10^{-3} &,& \sin^2(\theta_{13}) = 3.3\,10^{-5}.
\label{eq:nupred}
\end{eqnarray}
The most recent experimental bounds can be found in \cite{GGM}. They read
\begin{equation}
\sin^2 (\theta_{13}) \leq 0.05,
\end{equation}
which only leaves the smallest solution in (\ref{eq:nupred})
\footnote{Our predictions substantially differs from the ones in
\cite{Picariello}, which mainly focuses on special textures for 
the product of the quark and neutrino mixing matrices \cite{Xing}.}.

Future experiments will confirm, or infirm, for neutrinos,
the properties that
we have shown to be satisfied with an impressive accuracy
by quark mixing angles.

\section{Comments and open issues}
\label{section:questions}

\subsection{How close are mixing matrices to unitarity? Mixing angles and
mass spectrum}
\label{subsection:strength}

An important characteristic of the conditions that fix the mixing angles
is that they do not depend on the strength of the violation of the two
properties under scrutiny, namely, the absence of non-diagonal neutral
currents and the universality of the diagonal ones in the space of mass
eigenstates. Since only their ratio  is concerned, each
violation can be infinitesimally small.

This  is, on one side, fortunate since we have not yet been able to
calculate the magnitude of the violation of the unitarity of the
mixing matrices from, for example, mass ratios. 
The issue, for fundamental particles, turns indeed to be much more
difficult conceptually than it was for composite particles like neutral
kaons \cite{MaNoVi}. 

But, on the other side, this blurs the relation between the mixing
pattern and the fermionic mass spectrum
\footnote{A rigorous investigation of this connection was done in
\cite{MachetPetcov}. It however rests on the assumption (incorrect in QFT)
that a
system of coupled fermions can be described by a unique constant
mass matrix, which is diagonalized by a bi-unitary transformation. 
Then the so-defined ``fermion masses'' are not the
eigenvalues of the mass matrix, which makes all the more tricky the
connection with the poles of the full propagator in QFT.\label{ftn:diago}}.
This was already blatant with the emergence of maximal
mixing as a special set of solutions of the unitarization equations in
\cite{DuretMachet1}, which  did not depend of any special type of mass
hierarchy. The question now arises of finding, if any,
properties of the mass spectrum,
which are, through the products $K^\dagger K$ of mixing matrices,
compatible with an $SU(2)$ symmetric pattern of weak neutral currents.

\subsection{Which   mixing angles are measured}
\label{subsection:cc}

The results that have been exposed are valid for fermions of both electric
charges. They concern the mixing angles which parametrize

$\ast$ for quarks, the mixing matrix $K_u$
of $u$-type quarks as well as $K_d$ of d-type quarks;

$\ast$ for leptons, the mixing matrix $K_\nu$ of neutrinos as well as
that of charged leptons $K_\ell$,

and we have shown that our approach allows to obtain 
on purely theoretical grounds the values 
of the mixing angles which are experimentally determined.

However, a problem arises :  the measured values  of the mixing angles are
commonly attached, not to a single mixing matrix, {e.g.} $K_u$ or $K_d$,
but to the product  $K=K^\dagger_u K_d$  which occurs in charged
currents when both quark types are mass eigenstates.
Thus, in the standard approach, they are {\em a priori} related 
to an entanglement of the 
mixing angles of quarks (or leptons) of different charges.
This problem gets easily solved by the following argumentation.
Considering, for example,  semi-leptonic decays of pseudoscalar mesons
 in the approach where one of the constituent quarks is ``spectator'', we show
that only one of the two mixing matrices is involved and measured.
Indeed,  while the two-fold nature (flavor and mass)
of the neutral kaons has always been acknowledged, this
step has never been taken yet for other mesons. This is what we
do now, in a rather naive, but efficient way, that consists of
distinguishing a  $[\bar q_{i,f} q_{j,f}]$ ``flavor'' meson from the
mass eigenstate $[\bar q_{i,m} q_{j,m}]$ ($q_{i,j}$ being the
constituent quarks). Consider for example, the decay $K^0 \to \pi^- e^+
\nu_e$. The $K^0$ that decay semi-leptonically being produced
by strong interactions cannot be but a flavor meson $[\bar s_f d_f]$,
while its decay product $\pi^-$, which is identified by its mass and charge,
is a mass eigenstate $[\bar u_m d_m]$.
At the quark level, the weak transition occurs accordingly
between a flavour eigenstate ($\bar s_f$) to a mass eigenstate
($\bar u_m$), which only involves
{\em one} mixing matrix, $K_u$, and not the product
$K^\dagger_u K_d$.  As for the spectator quark, the transition from its
flavor state $d_f$ to its mass state $d_m$  involves the cosine of the
corresponding mixing angle, which is always close to $1$. It thus appears
that the mixing angles that are measured in such processes are the ones of
$K_u$ or $K_d$ (up to a cosine factor very close to $1$), which fits
with our symmetric prediction.

The same problem is expected in the leptonic sector.
Its solution depends on the nature of the neutrino eigenstates that are
produced and detected. Let us consider for example the case of solar
neutrinos.
If the flux predicted in solar models concerns  flavour neutrinos, and if
the detection process also counts flavour neutrinos, the sole mixing
matrix which controls their evolution and oscillations is $K_\nu$, because
it is the only matrix involved in the projection of flavour states onto mass
states.
This is the most likely situation. Indeed, the production mechanism inside
the sun occurs through nuclear beta decay, in which the protons and neutrons,
being bound by strong forces, are presumably, together with their
constituent quarks, flavour eigenstates. The detection (for example the
transition from chlorine to argon) also occurs  through nuclear (inverse)
beta decay, which accordingly also counts the number of $\nu_{ef}$ reaching the
detector.
The  situation would be different if the comparison was made between the
fluxes of the eigenstates $\nu_e, \nu_\mu, \nu_\tau$  defined in subsection
\ref{subsection:basis} (see also appendix \ref{section:CC}); since their
projections on the mass eigenstates involve the product $K_\ell^\dagger
K_\nu$, their oscillations are now controlled by an entanglement
of the mixing angles of neutrinos and charged leptons.

\subsection{A multiscale problem}
\label{subsection:mscale}

Recovery of the present results by
perturbative techniques (Feynman diagrams) stays an open issue.
All the subtlety of the problem lies in the inadequacy of using a single
constant mass matrix; because non-degenerate coupled systems are multiscale
systems, as many mass matrix should be introduced as there are poles in the
(matricial) propagator \cite{Novikov}
\footnote{In QFT, as opposed to a Quantum Mechanical
treatment (in which a single constant mass matrix is
introduced -- this is the Wigner-Weisskopf approximation--), a constant
mass matrix can only be introduced in a linear approximation to
the inverse propagator in the vicinity of each of its poles \cite{MaNoVi}.
When several coupled states are concerned, the (matricial) propagator
having several poles,  as many (constant) mass matrices should be
introduced \cite{Novikov};
only one of the eigenstates of each of these mass matrices
corresponds to a physical (mass) eigenstate.}.

The existence of different scales makes the use of an ``on-shell''
renormalized Lagrangian \cite{KniehlSirlin} hazardous,
because each possible renormalization
scale optimizes the calculation of parameters at this scale, while, for
other scales, one has to rely on renormalization group equations.

Unfortunately,  these equations
have only been approximately solved with the simplifying assumption that
the renormalized mass matrices are hermitian
\footnote{One can go to hermitian mass matrices by rotating right-handed
fermions {\em as far as they are not coupled}; however, at two loops, the
charged weak currents also involve right-handed fermions, which cannot be
anymore freely rotated.}
and that the renormalized mixing matrices are unitary \cite{KniehlSirlin}.
Performing the same job dropping these hypotheses looks rather formidable
and beyond the scope of the present work.
It also unfortunately turns out that, as far as the
Yukawa couplings are concerned, the expressions that have been obtained at
two loops for their $\beta$ functions (which start the evolution only up
 from the top quark mass) \cite{BHMP} have poles
in $(m_i -m_j)$, which makes them   inadequate for the study of
subsystems with masses below the top quark mass.

\subsection{Using a $\boldsymbol{q^2}$-dependent renormalized mass matrix}
\label{subsection:mamat}

Departure from the inappropriate Wigner-Weisskopf approximation
can also be done by  working
with an effective renormalized $q^2$-dependent mass matrix $M(q^2)$.
It however leads to similar conclusions as the present approach.

Its eigenvalues  are now $q^2$-dependent, and are determined by the
equation $\det[M(q^2) -\lambda(q^2)]=0$
\footnote{This is the simple  case of a normal mass
matrix,  which can be diagonalized by a single ($q^2$-dependent) unitary
matrix. When it is non-normal, the standard procedure uses a bi-unitary
diagonalization (see footnote \ref{ftn:diago}).}.
 Let them be $\lambda_1(q^2) \ldots
\lambda_n(q^2)$. The physical masses satisfy the $n$ self-consistent
equations $q^2 = \lambda_{1\ldots n}(q^2)$, such that
$m_1^2 = \lambda_1(m_1^2) \ldots m_n^2 = \lambda_n(m_n^2)$. At each
$m_i^2$, $M(m_i^2)$ has $n$ eigenvectors, but only one corresponds to the
physical mass eigenstate; the others are ``spurious'' states \cite{MaNoVi}.
Even if the renormalized mass matrix is hermitian at any given $q^2$,
the physical mass eigenstates corresponding to different $q^2$ belong to as
many different orthonormal sets of eigenstates and  thus, in general, do
not form an orthonormal set. The discussion proceeds like in the core of
the paper.

Determining the exact form of the renormalized mass matrix could
accordingly be a suitable way to recover our predictions via perturbative
techniques (like was done in \cite{MaNoVi} for the quantitative prediction
of the ratio  $\epsilon_S/\epsilon_L$).  As already mentioned, the difficulty is that
 hermiticity assumptions
should be dropped, which open the possibility of departing from the
unitarity of the mixing matrix. This is currently under investigation.

\section{Conclusion and perspective}
\label{section:conclusion}

This work does not, obviously, belong to what is nowadays referred 
to as "Beyond the Standard Model", since it does not incorporate any 
``new physics'' such as supersymmetry, ``grand unified theories (GUT)''
 or extra-dimensions.
However it does not strictly lie within
 the SM either, even if it is very close to.  Of course,
 it shares with the latter its general framework
(mathematical background and physical content), 
 and also borrows from it the two physical conditions of universality for 
 diagonal neutral currents  
 and absence of FCNC's, which play a crucial role in the process.
But, on the basis of the most general arguments of 
QFT, we make a decisive use of 
the essential non-unitarity of the mixing matrices,
whereas only unitary matrices are present in the SM.
This property  may be considered, in the SM,
as an "accidental" characteristic of objects which are
intrinsically non-unitary.

The mixing angles experimentally observed get constrained in the vicinity
of this ``standard'' situation, a slight departure from which
being due to  mass splittings.
Hence our approach can be considered to explore the
"Neighborhood of the Standard Model", which 
is likely to exhibit low-energy manifestations
of physics "Beyond the Standard Model".

While common approaches limit themselves to guessing symmetries
for the mass matrices
 (see for example \cite{Ma} and references therein),
we showed that special patterns are instead likely to reveal
themselves in the violation of some (wrongly) intuitive
properties
\footnote{
For a (constant unique) mass matrix, unitarity of the mixing matrix
has commonly been linked with the unitarity of the theory. See also
footnote \ref{ftn:unitarity}.}.
In each given $(i,j)$ channel of  mass eigenstates,
the characteristic pattern that emerges is that
two {\em  a priori} different  violations follow from a precise horizontal
continuous symmetry, which is the most intuitive $SU(2)$ group
attached to this pair of fermions. One simply falls back on an,
up to now unraveled, manifestation of  ``old Current Algebra''
\cite{GellMann}. It is remarkable that the same symmetry underlies both the
quark and leptonic sectors, which was never suspected before;
they only differ through the $0$th order
solution to the unitarization equations,
the two-foldness of which was recently uncovered in \cite{DuretMachet1}.
We have in particular learned that symmetries  relevant for flavour physics
should not be looked for, or implemented, at the level of the mass
matrices and Yukawa couplings,  but at the level of the weak currents.

We have also argued that, unlike what is generally assumed, the mixing
angles that are measured are (up to a cosine)
the ones of a single mixing matrix, and not of
the product $K^\dagger_u K_d$ or $K^\dagger_\ell K_\nu$.
Our scheme then appears
entirely coherent, and agrees with experimental data.

To conclude, the present work demonstrates that flavor physics 
satisfies very simple criteria which had been, up to now, unnoticed.
Strong arguments have been presented in both
the quark and leptonic sectors, which will be further tested when
the third mixing angle of neutrinos is accurately determined.


\vskip .5cm 
\begin{em}
\underline {Acknowledgments}: Discussions with A. Djouadi, J. Orloff and
 M.I. Vysotsky are gratefully acknowledged.
\end{em}


\newpage\null

\appendix

{\Large \bf Appendix}

\vskip 1cm

\section{$\boldsymbol{\tilde{\theta}_{13}=0 \Rightarrow \theta_{13}=0}$}
\label{section:theta3}

Using the notations of section \ref{section:general}, we start with
the following system of equations:
\begin{subequations}
\label{subeqs:A}
\begin{equation} \label{eq:I}
  \frac{[11]+[22]}{2}=[33] \Leftrightarrow
s_{13}^2+s_{23}^2+\tilde{c}_{23}^2=1;
\end{equation}
\begin{equation} \label{eq:IIA}
 [11]=[22] \Leftrightarrow  c_{13}^2\cos(2\theta_{12})=(c_{23}^2+
\tilde{s}_{23}^2)\cos(2\tilde{\theta}_{12});
\end{equation}
\begin{equation}\label{eq:IIB}
 [12]=0=[21] \Leftrightarrow 
c_{13}^2\sin(2\theta_{12})=(c_{23}^2+\tilde{s}_{23}^2)
\sin(2\tilde{\theta}_{12});
\end{equation}
\begin{equation}\label{eq:IIIA}
 [13]=0=[31] \Leftrightarrow
 \tilde{s}_{12}\left(\sin(2\theta_{23})-\sin(2\tilde{\theta}_{23})\right)
=c_{12}\sin(2\theta_{13});
\end{equation}
\begin{equation}\label{eq:IIIB}
 [23]=0=[32] \Leftrightarrow
\tilde{c}_{12}\left(\sin(2\tilde{\theta}_{23})-\sin(2\theta_{23})\right)
=s_{12}\sin(2\theta_{13}).
\end{equation}
\end{subequations}
From equation (\ref{eq:I}), we have 
$c_{23}^2+\tilde{s}_{23}^2 \neq 0$, 
which entails $c_{13}^2 \neq 0
$\footnote{Indeed, let us suppose that $c_{13}$ vanishes. 
Then $\cos(2\tilde{\theta}_{12})$ and $\sin(2\tilde{\theta}_{12})$ 
must vanish simultaneously, which is impossible.}.
Let us study the consequence on the two equations
(\ref{eq:IIA}) and (\ref{eq:IIB}).

$\bullet$ the two sides of (\ref{eq:IIA}) vanish for $\cos(2\theta_{12})=
0=\cos(2\tilde{\theta}_{12})$, {\em i.e.} 
$\theta_{12}=\frac{\pi}{4} [\frac{\pi}{2}]=\tilde{\theta}_{12}$.\\
(\ref{eq:IIB}) then gives  $c_{13}^2=c_{23}^2+\tilde{s}_{23}^2$,
which, associated with (\ref{eq:I}), yields the following solution
\footnote{$ \left\{\begin{array}{l}
c_{13}^2=c_{23}^2+\tilde{s}_{23}^2 \\ s_{13}^2+s_{23}^2+\tilde{c}_{23}^2=1 
\end{array}\right.
\qquad
 \Longrightarrow \qquad \left\{\begin{array}{l} s_{23}^2+\tilde{c}_{23}^2=1 \\
s_{13}^2=0
 \end{array}\right.$}: 
$\theta_{13}=0 [\pi]$ and $\tilde{\theta}_{23}=\pm \theta_{23} [\pi]$.

$\bullet$ the two sides of (\ref{eq:IIB}) vanish for $\sin(2\theta_{12})=
0=\sin(2\tilde{\theta}_{12})=0$, i.e. 
$\theta_{12}=0 [\frac{\pi}{2}]=\tilde{\theta}_{12}$.\\
 (\ref{eq:IIA}) gives then $c_{13}^2=c_{23}^2+\tilde{s}_{23}^2$, 
hence, like previously, $\theta_{13}=0 [\pi]$ and 
$\tilde{\theta}_{23}=\pm \theta_{23} [\pi]$.

$\bullet$ in the other cases we can calculate
 the ratio (\ref{eq:IIA}) / (\ref{eq:IIB}), which gives 
$\tan(2\theta_{12})=\tan(2\tilde{\theta}_{12})$, 
hence $\theta_{12}=\tilde{\theta}_{12} [\pi]$ or $\theta_{12}=
\frac{\pi}{2}+\tilde{\theta}_{12} [\pi]$:
 
\quad$\ast$ $\theta_{12}=\frac{\pi}{2}+\tilde{\theta}_{12} [\pi]$ implies 
for (\ref{eq:IIA})(\ref{eq:IIB}) $c_{13}^2=-c_{23}^2-\tilde{s}_{23}^2$, 
which, together with (\ref{eq:I}) ($c_{13}^2=s_{23}^2+\tilde{c}_{23}^2$),
gives a contradiction : $2=0$:

\quad$\ast$ $\theta_{12}=\tilde{\theta}_{12} (\neq 0)[\pi]$ implies, like 
previously, $c_{13}^2=c_{23}^2+\tilde{s}_{23}^2$, 
which gives, when combined with (\ref{eq:I}):
$\theta_{13}=0 [\pi]$ and $\tilde{\theta}_{23}=\pm \theta_{23} [\pi]$.

Hence, it appears that whatever the case, the solution gives rise
to $\theta_{13}=0 [\pi]$.

Let us now look at (\ref{eq:IIIA}) and (\ref{eq:IIIB}).
Since $\theta_{13}=0$, the two r.h.s.'s vanish, 
and we obtain 
the twin equations $\tilde{s}_{12}(\sin(2\theta_{23})-\sin(2\tilde{\theta}_{23}))=0$ and 
 $\tilde{c}_{12}(\sin(2\theta_{23})-\sin(2\tilde{\theta}_{23}))=0$,
which, together, imply 
 $\sin(2\theta_{23})=\sin(2\tilde{\theta}_{23})$. It follows that, either 
 $\theta_{23}=\tilde{\theta}_{23} [\pi]$ or 
 $\theta_{23}=\frac{\pi}{2}-\tilde{\theta}_{23} [\pi]$;
 
\quad$\ast$ $\theta_{23}=\tilde{\theta}_{23} [\pi]$ matches the result of the 
previous discussion in the ``+" case, whereas, in the ``-" case,
 the matching leads to 
$\theta_{23}=\tilde{\theta}_{23}=0$, which is to be absorbed
as a particular case in the ``+" configuration;
 
\quad$\ast$ $\theta_{23}=\frac{\pi}{2}-\tilde{\theta}_{23} [\pi]$ matches 
the result of the previous discussion in the ``+" configuration,
in which case it leads to 
$\theta_{23}=\tilde{\theta}_{23}=\frac{\pi}{4} [\frac{\pi}{2}]$, {\em i.e.}
maximal mixing 
between the fermions of the second and third generations.

\section{$\boldsymbol{(\theta_{12},\theta_{23})}$ solutions of eqs.
(\ref{eq:nodb}) (\ref{eq:nosb}) (\ref{eq:nods}) (\ref{eq:ddss})
(\ref{eq:ssbb}) for
$\boldsymbol{\theta_{13} = 0 = \tilde{\theta}_{13}}$}
\label{section:sol0}

Excluding $\tilde{\theta}_{12}=0$, (\ref{eq:nodb0}) and (\ref{eq:nosb0}) require
$\sin(2\theta_{23}) = \sin(2\tilde{\theta}_{23}) \Rightarrow \tilde{\theta}_{23} = \theta_{23} + k\pi$
 or $\tilde{\theta}_{23} = \pi/2 - \theta_{23} + k\pi$.

$\bullet$ for $\tilde{\theta}_{23} = \theta_{23} + k\pi$ Cabibbo-like,

(\ref{eq:nods0}) requires
$\sin(2\theta_{12}) = \sin(2\tilde{\theta}_{12})  \Rightarrow
\tilde{\theta}_{12} = \theta_{12} + n\pi$ or $\tilde{\theta}_{12} = \pi/2 - \theta_{12} + n\pi$;

(\ref{eq:ddss0}) requires
$\cos(2\theta_{12}) = \cos(2\tilde{\theta}_{12}) \Rightarrow \tilde{\theta}_{12} = \pm \theta_{12} +
p\pi$;

(\ref{eq:ssbb0}) requires $s_{12}^2 + \tilde c_{12}^2 -1 =0 \Rightarrow
\tilde{\theta}_{12} = \pm \theta_{12} + r\pi$.

The solutions of these three equations are
$\theta_{12} = \tilde{\theta}_{12} + k\pi$ Cabibbo-like or
$\theta_{12} = \pi/4 + q\pi/2$ maximal.

$\bullet$ for $\tilde{\theta}_{23} = \pi/2 - \theta_{23} + k\pi$,

(\ref{eq:nods0}) requires $s_{12}c_{12} = 2c_{23}^2 \tilde s_{12} \tilde
c_{12}$;

(\ref{eq:ddss0}) requires $c_{12}^2 -s_{12}^2 = 2 c_{23}^2 (\tilde c_{12}^2 - \tilde
s_{12}^2)$;

(\ref{eq:ssbb0}) requires $s_{12}^2 + 2c_{23}^2 \tilde c_{12}^2 - 2s_{23}^2
=0$.

The first two conditions yield $\tan(2\theta_{12}) = \tan(2\tilde{\theta}_{12})
 = 2c_{23}^2
\Rightarrow \tilde{\theta}_{12} = \theta_{12} + k\pi/2 + n\pi$, which entails
$2c_{23}^2 = 1 \Rightarrow \theta_{23} = \pm \pi/4 + p\pi/2$ maximal; $\tilde{\theta}_{23}$
is then maximal, too, and the third
condition is automatically satisfied.

$\tilde{\theta}_{12} = \theta_{12} + n\pi$ is Cabibbo-like, while, for 
$\tilde{\theta}_{12} =
\theta_{12} + (2k+1)\pi/2$, the second condition becomes $(c_{12}^2 -s_{12}^2)=0$,
which means that $\theta_{12}$ must be maximal.

\section{Sensitivity of the neutrino solution to a small variation of
$\boldsymbol{\theta_{13}}$}
\label{section:maxsens}
 
If one allows for a small $\theta_{13} \approx \tilde{\theta}_{13}$, 
(\ref{eq:nods}) and
(\ref{eq:ddss}) become
\begin{eqnarray}
&& -2\eta  s_{12} c_{12} s_{23} c_{23} + \epsilon (s_{12}^2 - c_{12}^2) 
            +\eta s_{13} (c_{23}^2 -s_{23}^2)(c_{12}^2 - s_{12}^2), \cr
&& -2\eta s_{23}c_{23} (c_{12}^2 - s_{12}^2) + 4 \epsilon s_{12} c_{12} 
     -2\eta s_{13} (c_{23}^2 - s_{23}^2)(2s_{12}c_{12} + \epsilon(c_{12}^2 - s_{12}^2)).
\end{eqnarray}
For $\theta_{23}, \tilde{\theta}_{23}$ maximal, the dependence on $\theta_{13}$ drops out.

\section{Charged weak currents}
\label{section:CC}

Charged weak currents can be written in six different forms that are all 
strictly equivalent, but nonetheless refer to different physical pictures. As an 
example, for two generations of leptons :
\begin{eqnarray}
&&\overline{\left(\begin{array}{c} \nu_{ef} \cr \nu_{\mu f}\end{array}
\right)}
W_\mu^+  \gamma^\mu_L
\left(\begin{array}{c} e^-_f \cr \mu^-_f\end{array} \right)
=
\overline{\left(\begin{array}{c} \nu_{em} \cr \nu_{\mu m}\end{array} \right
)}
W_\mu^+  \gamma^\mu_L K^\dagger_\nu K_\ell
\left(\begin{array}{c} e^-_m \cr \mu^-_m\end{array} \right)\cr
&=&
\overline{\left(\begin{array}{c} \nu_{ef} \cr \nu_{\mu f}\end{array} \right
)}
W_\mu^+  \gamma^\mu_L \left[K_\ell
\left(\begin{array}{c} e^-_m \cr \mu^-_m\end{array} \right)\right]
=
\overline{\left[K_\nu
\left(\begin{array}{c} \nu_{em} \cr \nu_{\mu m}\end{array}\right)
\right]} W_\mu^+  \gamma^\mu_L
\left(\begin{array}{c} e^-_f \cr \mu^-_f\end{array} \right)\cr
&=&
\overline{\left(\begin{array}{c} \nu_{em} \cr \nu_{\mu m}\end{array} \right
)}
W_\mu^+  \gamma^\mu_L \left[K^\dagger_\nu
\left(\begin{array}{c} e^-_f \cr \mu^-_f\end{array} \right)\right]
=
\overline{\left[K^\dagger_\ell\left(\begin{array}{c} \nu_{ef} \cr \nu_{\mu
f}\end{array}
\right)\right]}
W_\mu^+  \gamma^\mu_L
\left(\begin{array}{c} e^-_m \cr \mu^-_m\end{array} \right).
\label{eq:couplings}
\end{eqnarray}
In the case where one of the $SU(2)$ partners, for example the charged lepton,
 is undoubtedly a mass eigenstate
\footnote{This is the case inside the sun \cite{Vysotsky}
 where, because of the limited
available energy, only massive electrons can be produced, and also in the
detection process of neutrinos on earth, which always proceeds via charged
currents and the detection of produced physical (mass eigenstates)
charged leptons.}
, the last expression of
(\ref{eq:couplings}) shows that it is coupled to the so-called
{\em  electronic and muonic neutrinos}
\begin{eqnarray}
\nu_e = (K^\dagger_\ell K_\nu)_{11} \nu_{em} + (K^\dagger_\ell K_\nu)_{12}
\nu_{\mu m} &=& K^\dagger_{\ell,11} \nu_{ef} + K^\dagger_{\ell,12} \nu_{\mu
f},\cr
\nu_\mu = (K^\dagger_\ell K_\nu)_{21} \nu_{em} + (K^\dagger_\ell K_\nu)_{22}
\nu_{\mu m} &=& K^\dagger_{\ell,21} \nu_{ef} + K^\dagger_{\ell,22} \nu_{\mu f}.
\label{eq:elecnu2}
\end{eqnarray}
The latter are neither flavour eigenstates, nor mass eigenstates, but
a third kind of neutrinos, precisely defined as the ones
which couple to electron and muon mass eigenstates in the weak
charged currents
\begin{equation}
\overline{\left(\begin{array}{c} \nu_{e} \cr \nu_{\mu}\end{array} \right)}
W_\mu^+  \gamma^\mu_L
\left(\begin{array}{c} e^-_m \cr \mu^-_m\end{array} \right).
\label{eq:nuenumu}
\end{equation}
%

\newpage\null
\begin{em}

\end{em}

\end{document}